\documentclass[12pt]{article}
\usepackage{amssymb,amsmath}
\include{epsf}

\usepackage{graphicx}
\usepackage{amsfonts}

\newcommand{\e}{{\mathrm e}}
\newcommand{\ii}{{\mathrm i}}
\newcommand{\dd}{{\mathrm d}}
\newcommand{\eqn}[1]{(\ref{#1})}

\def\appendix#1{\addtocounter{section}{1}\setcounter{equation}{0}
\renewcommand{\thesection}{\Alph{section}}
\section*{
\thesection\protect\indent \parbox[t]{11.715cm} {#1}}
\addcontentsline{toc}{section}{Appendix\thesection\ \ \ #1} }
\newcommand{\real}{{\mathbb R}} 

\newcommand{\be}{\begin{equation}}
\newcommand{\ee}{\end{equation}}
\newcommand{\beq}{\begin{equation}}
\newcommand{\eeq}{\end{equation}}
\newcommand{\bea}{\begin{eqnarray}}
\newcommand{\eea}{\end{eqnarray}}
\def\beqa{\begin{eqnarray}}
\def\eeqa{\end{eqnarray}}
\def\nn{\nonumber}
\newcommand{\del}{\partial}

\newcommand{\eq}{\begin{equation}}
\newcommand{\eqa}{\begin{eqnarray}}
\newcommand{\en}{\end{equation}}
\newcommand{\ena}{\end{eqnarray}}

\newcommand{\starmoy}{\star_M}
\newcommand{\starvor}{\star_V}

\begin{document}
\begin{titlepage}
\begin{flushright}

\baselineskip=12pt
DSF--12--2009\\ ICCUB-09-219\\July 2009\\
\hfill{ }\\
\end{flushright}

\begin{center}

\baselineskip=24pt

{\Large\bf
Translation Invariance,\\ Commutation Relations and Ultraviolet/Infrared Mixing}

\baselineskip=14pt

\vspace{1cm}

{\bf Salvatore Galluccio$^1$, Fedele Lizzi$^{1,2}$ and Patrizia Vitale$^1$}
\\[6mm]
$^1${\it Dipartimento di Scienze Fisiche, Universit\`{a}
di Napoli {\sl Federico II}}\\ and {\it INFN, Sezione di Napoli}\\
{\it Monte S.~Angelo, Via Cintia, 80126 Napoli, Italy}
\\[4mm]
$^2$ {\it Institut de Ci\'encies del Cosmos,
Universitat de Barcelona,\\
Marti i Franqu\`es 1, Barcelona, Catalonia, Spain}\\{\small\tt
salvatore.galluccio@na.infn.it, fedele.lizzi@na.infn.it,
patrizia.vitale@na.infn.it}

\end{center}

\hfill \textit{Dedicated to the memory of Raffaele Punzi}

\vskip 2 cm

\begin{abstract}
We show that the Ultraviolet/Infrared mixing of noncommutative
field theories with the Gr\"onewold-Moyal product, whereby some
(but not all) ultraviolet divergences become infrared, is a
generic feature of translationally invariant associative products.
We find, with an explicit calculation that the phase appearing in
the nonplanar diagrams is the one given by the commutator of the
coordinates, the semiclassical Poisson structure of the non
commutative spacetime. We do this with an explicit calculation for
represented generic products.
\end{abstract}

\end{titlepage}

\section{Introduction}
\setcounter{equation}{0}

One of the original motivations~\cite{Heisenberg, Snyder} to consider a noncommutative structure of space or spacetime was the hope that the presence of a dimensionful parameter, and a modification of the short distance properties, could resolve the problem of the infinities of quantum field theory. The analogy in this case is the presence of $\hbar$ and the noncommutativity of qantum phase space solves the so called ultraviolet catastrophe of the black body radiation. In the case of a field theory described by the Gr\"onewold-Moyal product this hope is not fulfilled. In this case instead of the elimination (at least partial) of the ultraviloet infinities, we enconter the phenomenon of ultraviolet/infrared mixing~\cite{MvRS}, one of the novel features of a field theory over a noncommutative space (noncommutative field theory).

Technically this means that some ultraviolet divergences of the
ordinary theory disappear, at the price of the appearance of
infrared divergences in the same diagrams. In particular one finds
that this happens at one loop for nonplanar diagrams. Therefore
while the ultraviolet, short distance, properties of the theory
are changed in the sense of a mitigation of the infinities, the
price paid is the appearance of new kind of infinity. We will
describe in detail this phenomenon for the one loop corrections to
the propagator, but there is also a rough heuristic explanation of
this phenomenon. The noncommutative $\star$ product used in the
Gr\"onewold-Moyal product reproduces the commutation relation of
quantum mechanics: $[x^i,x^j]=\ii\theta^{ij}$. Coordinates do not
commute and therefore a generalization of Heisenberg's uncertainty
principle is at work, a small uncertainty in the $x^i$ direction
implies a great uncertainty in the $\theta^{ij}x_j$ direction.
Therefore a short distance in one direction and the long distance
in the other are coupled. This reasoning can be made more
precise~\cite{MvRS} (but still heuristic) considering the
dispersion of the product of gaussian functions. The phenomenon
persists also in the nonrelativistic
case~\cite{GomisLandsteinerLopez}.

The aim of this paper is to discuss the ultraviolet structure of
noncommutative theories with more general products than
Gr\"onewold-Moyal. Our analysis will be centered on the one loop
correction to the propagator, which is the source of all mixing.
We will discuss only the bosonic $\phi^4$ theory, but the results
are more general than that and will apply to other scalar and
gauge theories as well since, as we will see, the behaviour which
we find is quite generic.

We will be discussing the euclidean version of the theory, or
equivalently the case of only spatial commutativity. In the
Minkowskian case the nonlocality of the theory has been claimed to
lead to loss of unitarity~\cite{GomisMehen} in the noncommutative
theory which is obtained as an  effective theory of
strings~\cite{SeibergWitten}. Nevertheless for theories for which
noncommutativity is fundamental there are issues of time
ordering~\cite{BahnsDoplicherFredenhagenPiacitelli,LiaoSibold1,
LiaoSibold2, BFGPPSW} which show that an appropriate treatment can
lead to an unitary theory. For purely time-space noncommutativity
the mixing may not present as such~\cite{Bahns}.

The ultraviolet/infrared mixing is in general connected with the
nonlocality of the product and has been generalised in various
directions. Gayral~\cite{Gayral} has shown that it persists in the
presence of isospectral deformations. The
noncommutativity for the compact case is basically given by a
noncommutative torus, which in this context is a compact version
of the Gr\"onewold-Moyal product. Some form of mixing also
survives for the $\kappa$-Minkowski case~\cite{GrosseWohlgenannt}.


The paper is organized as follows. In section 2 we discuss the
Ultraviolet/Infrared Mixing for the Gr\"onewold-Moyal Product. We
then discuss the general form of translationally invariant
products. In section 4 we show the form of the mixing for a
general product. This section contains the main result of the
paper, that is that the mixing persists unchanged for a generic
translation invariant product. We end the paper with some conclusions.

\section{Ultraviolet/Infrared Mixing for the\\ Gr\"onewold-Moyal Product}
\setcounter{equation}{0}
In this section we review the presence of ultraviolet/infrared mixing for a scalar theory. We consider a field theory on a noncommutative space described by the action:
\be
S=\int \dd x^d \frac12\left(\del_i\varphi\star\del_i\varphi
-m^2\varphi\star\varphi\right)+\frac{g}{4!}
\varphi\star\varphi\star\varphi\star\varphi \label{action4}
\ee
where $\star$ usually denotes the Gr\"onewold-Moyal product between functions which can be defined in several different ways. These definitions are equivalent up to the fact that the domain of definition can be different. The product depends on an antisymmetric matrix $\theta^{ij}$ and we write two standard expressions of it. The most common expression is expressed as a series of differential operators:
\be
\left.(f\starmoy g)(x)=\e^{\frac{\ii}{2}\theta^{ij}\del_{y_i}\del_{z_j}}f(y)g(z)\right\vert_{x=y=z}
\label{Moyalseries}
\ee
This series is an asymptotic expansion~\cite{EstradaGraciaBondiaVarilly} of (equivalent) integral expressions, some of which can be found in the appendix of~\cite{Selene}. For the purposes of this paper the useful form of the product is the following:
\be
(f\starmoy g)(x)=\frac1{(2\pi)^{\frac d2}} \int \dd^d p \, \dd^d q
\tilde f(q) \tilde g(p-q) \e^{\ii p\cdot x} \e^{\frac{\ii}{2}
p_i\theta^{ij} q_j} \label{MoyalFourier}
\ee
where $\tilde f$ and $\tilde g$ are the usual Fourier transforms of $f$ and $g$ respectively. In both cases it results
\be
[x^i,x^j]_{\starmoy}=\ii\theta^{ij} \label{xcommrel}
\ee
and the product becomes the ordinary, commutative product for $\theta=0$. Note that for this product
\be
\int\dd x^d f\starmoy g=\int\dd x^d f g \label{intMoyal}
\ee
which means that the free (quadratic) theory is the same in the commutative and noncommutative cases.

\begin{figure}[htbp]
\epsfxsize=4.5 in
\bigskip
\centerline{\epsffile{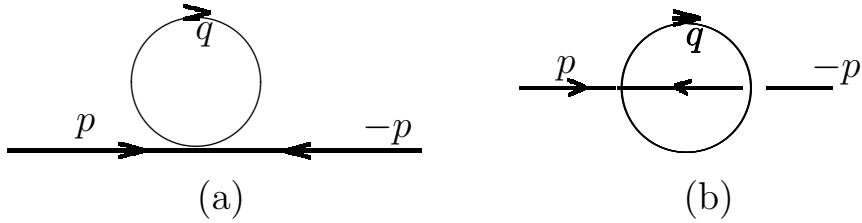}}
\caption{\sl The planar (a) and nonplanar (b) one loop correction
to the propagator} \label{oneloop}
\end{figure}

The theory described by the action~\eqn{action4} has a propagator which is the same as in the commutative case and a vertex~\cite{Filk} which is easily calculated from~\eqn{MoyalFourier} to be, for four incoming propagators of momenta~$k_a$,
\be
V_{\mathrm{Moyal}}=V_0\e^{-\frac\ii2\sum_{a\leq b}\theta^{ij}{k_a}_i{k_b}_j}
\ee
where
\be
V_0=-\ii \frac{g}{4!} (2\pi)^d\delta^d\left(\sum_{a=1}^4 \label{V0def}
{k_a}\right)
\ee
is the usual vertex of the commutative theory. The new vertex is not anymore invariant for the exchange of incoming propagators, but maintains invariance for cyclic permutations. As a consequence the planar and nonplanar diagrams are not necessarily equal and have to be calculated separately. In this paper we will limit ourselves to the one loop case because we are interested in the generic behaviour in the ultraviolet. Therefore we will be looking at the two diagrams  described in Fig.~\ref{oneloop}
and the one loop corrections to the propagator. The corresponding Green's functions are
\bea
G^{(2)}_{\mathrm P} &=& - \ii \frac{g}{3}
\int\frac{\dd q^d}{(2\pi)^d}\frac{1}{(p^2-m^2)^2(q^2-m^2)}\nonumber\\ G^{(2)}_{\mathrm NP} &=& - \ii
\frac{g}{6}
\int\frac{\dd q^d}{(2\pi)^d}\frac{\e^{\ii p_i \theta^{ij}q_j}}{(p^2-m^2)^2(q^2-m^2)} \label{Moyalnonplanar}
\eea
In particular we see that the planar diagram is the same as in the commutative case, thus dashing the hope that this particular noncommutative theory, with its inherent nonlocality, could solve the infinities of field theory. The persistence of some divergences is more general than the present calculation and was noted in~\cite{varillygraciabondia} in the general framework of Connes' noncommutative geometry~\cite{Connesbook}, while in~\cite{ChaichianDemichevPresnajder} it is shown that not all divergences can be eliminated in the presence of the commutation relation~\eqn{xcommrel}.

Let us concentrate on the nonplanar diagram. For this diagram there are no ultraviolet divergences, and it is this diagram that shows the ultraviolet/infrared mixing. For high momentum $p$ the phase in the numerator oscillates rapidly and renders the diagram convergent. However the numerator vanishes for $p\to 0$ and we have
\be
\lim_{p_i\theta^{ij}\to 0}G^{(2)}_{\mathrm NP}=\frac12G^{(2)}_{\mathrm P}
\ee

In~\cite{paganini} following the procedure set in~\cite{Mozart} we have shown that the ultraviolet/infrared mixing persists in an unchanged way also for a variant of the Gr\"onewold-Moyal product, the Wick-Voros product. This is naturally defined in two dimensions but can be generalized to higher dimensions. Define:
\be
z_\pm=\frac{x^1\pm\ii x^2}{\sqrt{2}}  \label{defz}
\ee
We will also use the notation
\be
k_\pm=\frac{k_1\pm\ii k_2}{\sqrt{2}}
\ee
for a generic vector $\vec k$.

The series form of the
Wick-Voros product, analog of~\eqn{Moyalseries} is
\be
f\starvor g=\sum_n \left(\frac{\theta^n}{n!}\right) \del_+^n f
\del_-^n g = f
\e^{\theta\overleftarrow{\del_+}\overrightarrow{\del-}}g\label{vorosprod}
\ee
where
\be
\del_\pm=\frac{\del}{\del
z_\pm}=\frac{1}{\sqrt2}\left(\frac\del{\del
{x^1}}\mp\ii\frac\del{\del {x^2}}\right)
\ee
The integral expression analog of~\eqn{MoyalFourier} is
\be
(f\starvor g)(x)=\frac1{(2\pi)^{\frac d2}} \int \dd^d p \, \dd^d q
\tilde f(q) \tilde g(p-q) \e^{\ii p\cdot x} \e^{- \theta
q_-(p_+-q_+)} \label{WVFourier}
\ee
It results
\bea
z_+\starvor  z_-&=& z_+z_- +\theta\nonumber\\
z_-\starvor  z_+ &=& z_+z_-
\eea
and therefore
\be
[z_+,z_-]_{\starvor}=\theta
\ee
Going back to the $x$'s, it is possible to see that this relation
gives rise again to the standard commutator among the $x$'s:
\be
x^1\starvor x^2- x^2 \starvor x^1=\ii \theta
\ee

With the $z_\pm$  coordinates the Laplacian and the d'Alembertian
are respectively $\nabla^2=2 \del_+\del_-$ and
$\Box=\del_0^2-\nabla^2$. The integral on the plane is still a
trace, but the strong condition of~\eqn{intMoyal} is not valid
anymore:
\be
\int \dd^2 z  f\starvor g=\int \dd^2 z  g\starvor f\neq \int \dd^2
z f  g
\ee
where by $\dd^2z$ we mean the usual measure on the plane $\dd z_+\dd
z_-$.  This means that the free propagator is not the same anymore as it receives a correction\footnote{In the following, for this subsection, we will be in 2+1 dimensions to ease the comparison with the Moyal product. The results are more general and dimension independent.} by a factor $\e^{-\frac\theta2|\vec p|^2}$. The vertex has been calculated~\cite{paganini} and is
\be
V_{\starvor}=V\prod_{a<b}\e^{-\theta{{k_a}_-}{{k_b}_+}}=V\prod_{a<b}\e^{-\frac{\theta}{2}(
{ \vec {k_a}}\cdot{\vec{k_b}}+i \varepsilon^{ij}{k_a}_i{k_b}_j)}
\label{vertexVoros}
\ee
It is then possible to calculate the one loop correction to the propagator. For the planar case we obtain
\bea
G^{(2)}_{\mathrm P} &=& - \ii
\frac{g}{3}\int\frac{\dd^3q}{(2\pi)^3}\frac{\e^{-\theta(2p_-p_+
+q_- q_+)}\e^{-\theta(p_- q_+ - p_- q_+ - p_- p_+ -q_- q_+ -q_-
p_+ +q_- p_+)}}{(p^2-m^2)^2(q^2-m^2)}
\nonumber\\
&=&  - \ii \frac{g}{3}
\int\frac{\dd^3q}{(2\pi)^3}\frac{\e^{-\theta p_- p_+
}}{(p^2-m^2)^2(q^2-m^2)}
\eea
In this case all the contribution due to $q$
cancel, so that there is no change in the convergence of the
integral. This is the same as in the Gr\"onewold-Moyal case. The only difference with is again the correction to the propagator.
The expression for the nonplanar case is:
\bea
G^{(2)}_{\mathrm NP} &=& - \ii \frac{g}{6}
\int\frac{\dd^3q}{(2\pi)^3}\frac{\e^{-\theta(2p_- p_+ +q_-
q_+)}\e^{-\theta(p_- q_+ -p_- p_+ - p_- q_+ -q_- p_+ -q_- q_+ +
p_- q_+)}}{(p^2-m^2)^2(q^2-m^2)}
\nonumber\\
&=& - \ii
\frac{g}{6}\int\frac{\dd^3q}{(2\pi)^3}\frac{\e^{-\theta(p_- p_+
+\ii \vec p\wedge \vec q)}}{(p^2-m^2)^2(q^2-m^2)}
\eea
This time the $q$ contribution does not cancel completely, and
there remains the exponential of the factor
\be
p_- q_+ - q_- p_+= \ii \vec p \wedge \vec q
\ee
which is the same as in the Moyal case. We see that the ultraviolet behaviour of the two products is the same.
The presence of the term $\e^{\theta p^2}$ is due to the fact that the free propagator is different in this case from the commutative theory, which in turn is a consequence of the fact that for the Wick-Voros product the property~\eqn{intMoyal} does not hold, but the integral is still tracial ($\int\dd x^d f\star g=\int\dd x^d  g\star f$). Apart from this difference the structure is the same in the two theories, namely the one loop diagram does not give extra contributions in the planar case, while it does in the non planar one. We will see below that this behaviour is generic for all translation invariant products.

\section{General Translation Invariant Products}
\setcounter{equation}{0}
In this section we first introduce a generic star product in the differential series form, and in the integral form. General star products were introduced in~\cite{BFFLS1,BFFLS2} in the framework of quantization of Poisson manifolds. For our purposes a generic star product is an associative product between functions on $\real^d$ which depends on one or more parameters. In the limit in which these parameters vanish the product becomes the usual poinwise product. Notice that we contemplate the possibility that the star product be commutative, although in general it will not be so.

We will consider two ways of expressing these generic products. In most cases (as in the Gr\"onewold-Moyal case) these two ways coincide on a dense domain on some space of functions.  The problem with expressions like~\eqn{starexpans} is that they are defined only at the level of formal series, and there is no certainty that one can actually find a representation of the deformed algebra of (a class of) the functions on spacetime with the noncommutative product they define. We prefer to adhere to the principle: \emph{no deformation without representation}~\cite{Piacitelli} and will present first the integral form of the product, which is more suited for our purposes. Later we will present the generic differential form as well, and will comment throughout the paper on both forms  of the product.

The generalization of the star product \eqn{MoyalFourier} (or the
Wick-Voros product \eqn{WVFourier}) is the following
\be
f\star g=\frac1{(2\pi)^{\frac d2}}\int\dd p^d \dd q^d \dd k^d \e^{\ii p \cdot x} \tilde f(q)\tilde g(k) K(p,q,k) \label{intprod}
\ee
Where $K$ can be a distribution and $\tilde f(q)$ is the Fourier tranform of $f$. The product of $d$-vectors is understood with the Minkowski or Euclidean metric: $p\cdot x=p_ix^i$.
The usual pointwise product is also of this kind for $K(p,q,k)=\delta^d(k-p+q)$. The biggest restrictions on $K$ come from the associativity requirement which reads
\be
\int\dd k^d K(p,k,q)K(k,r,s)=\int\dd k^d K(p,r,k)K(k,s,q)
\label{cocyclecondition}
\ee
This is nothing but the usual cocycle condition in the Hochschild cohomology, where the two cocycle $c\in C^2 (\mathcal{A})$ is the map
\bea
 c: (f,g)&\in&\mathcal{A}\otimes\mathcal{A}\longrightarrow\mathcal{A}\nn\\
 c(f,g)&=&f\star g \label{coc}
\eea
$\mathcal{A}$ is the noncommutative algebra of functions with the star-product \eqn{intprod} and the coboundary operator
\be
\del: C^k (\mathcal{A}) \longrightarrow C^{k+1} (\mathcal{A})
\ee
\bea
\del c(f_0, ..., f_{k})&=& f_0\star c(f_1,...,f_k)+ \sum_{i=0}^{k-1}(-1)^{i+1}c(f_0,...,f_i\star f_{i+1},...,f_k)\nn\\
&+&(-1)^{k+1}c((f_0, ..., f_{k-1})\star f_k.
\eea
In order for the two-cochain \eqn{coc} to be a  two-cocycle this becomes
\bea
0= \del c(f,g,h)&=&f\star c(g,h) - c(f\star g,h) + c(f,g\star h) -c(f,g)\star h\nn\\
&=& 2\left(f\star(g\star h) -(f\star g) \star h\right)
\eea
that is \eqn{cocyclecondition}.

\newcommand{\tran}{\mathcal T}

We now proceed to the discussion on   translation invariant products. Defining the translation by a vector $a$ by $\tran_a(f)(x)=f(x+a)$, by translation invariant product we mean the property
\be
\tran_a(f)\star \tran_a(g)=\tran_a(f\star g)
\ee
At the level of Fourier transform we have
\be
\widetilde{\tran_a f}(q)=\e^{\ii a p}\tilde f(q)
\ee
For the invariance of the product~\eqn{intprod} we must have
\bea
\e^{\ii a\cdot p} \int \dd p^d \dd q^d \dd k^d \e^{\ii p \cdot x} \tilde f(q)\tilde g(k) K(p,q,k)=\nonumber\\
= \int \dd q^d \dd k^d \e^{\ii a\cdot q} \e^{\ii a\cdot k}\e^{\ii p \cdot x} \tilde f(q)\tilde g(k) K(p,q,k)
\eea
which means that at the distributional level
\be
K(p,q,k)=\e^{\ii(k-p+q)\cdot a}K(p,q,k)
\ee
which is solved by
\be
K(p,q,k)=\e^{\alpha(p,q)}\delta(k-p+q) \label{tinv}
\ee
where $\alpha$ is a generic function. We will therefore consider products that can be expressed as
\be
f\star g=\frac1{(2\pi)^{\frac d2}}\int\dd p^d \dd q^d  \e^{\ii p \cdot x} \tilde f(q)\tilde g(p-q) \e^{\alpha(p,q)} \label{intprodtran}
\ee
The usual pointwise product is given by $\alpha=0$, the Gr\"onewold-Moyal product by $\alpha_M (p,q)=-\ii/2\theta^{ij}q_ip_j$ and the Wick-Voros by $\alpha_V(p,q)=-\theta q_-(p_+-q_+)$.

Associativity and the requirement that the integral is a trace
impose severe constraints on the form of $\alpha$. In particular from
 \eqn{cocyclecondition} and  \eqn{tinv} the cocycle condition becomes
\be
\alpha(p,q)+\alpha(q,r)=\alpha(p,r)+\alpha(p-r,q-r) \label{associativity}
\ee
from this cocycle relation follow some other useful relations:
\bea
\alpha(p,p)&=&\alpha(0,0)=\alpha(p,0)\nonumber\\
\alpha(0,p)&=&\alpha(0,-p)\nonumber\\
\alpha(p,q)&=&-\alpha(q,p)+\alpha(0,q-p)\nonumber\\
\alpha(p+q,p)&=&-\alpha(0,p+q)+\alpha(0,p)+\alpha(0,q)-\alpha(-p-q,-q)
\label{relations}
\eea
This last relation ensures also the trace property.
\bea
\int\dd x^d f \star  g&=&\int\dd x^d \dd p^d \dd q^d \e^{\alpha(p,q)} \e^{\ii p\cdot x} \tilde f(q) \tilde g (p-q)\nonumber\\
&=&\int \dd q^d \e^{\alpha(0,q)} \tilde f(q) \tilde g (-q)
\eea
Another  relation which will be useful in the following is
\begin{equation}\label{IR}
\alpha(p,q)=-\alpha(0,p)+\alpha(0,q)+\alpha(0,p-q)-\alpha(-p,q-p).
\end{equation}
We also require the algebra to be a $*$-algebra. That is that
there is a $*$ conjugation such that ${f^*}^*=f$ and $(f\star
g)^*=g^*\star f^*$. This latter relation imposes
\be
\alpha(p,q)^*=\alpha(-p,q-p) \label{starcondition}
\ee
Note that we do not require necessarily $f\star 1=1\star f=f$, that is that the identity of the algebra is the constant function. This condition would impose
\begin{equation}
\alpha(p,p)=0\quad\mathrm{and}\quad\alpha(p,0)=0 \label{identitycondition}
\end{equation}

The $\star$ products that we are considering are in general
noncommutative, but a product of the form~\eqn{intprodtran} can be
commutative. In this case we have that the restriction on the kernel $K$
reads
\be
K(p,k,q)= K(p,q,k)
\ee
that is the cocycle $c$ is a coboundary
\be
c(f,g)= \del b(f,g) = f\star b(g) + g\star b(f) - b(f\star g) \label{coboundary}
\ee
with the cochain $b$ given by the identity map.
In terms of
$\alpha$ the coboundary condition becomes
\be
\alpha(p,q)=\alpha(p,p-q) \label{commutativeprod}
\ee
\subsection{Cohomology}
It is possible to define an``$\alpha$-cohomology'' with respect to which $\alpha$ is a 2-cocycle, while it becomes a coboundary for a commutative product. $ \alpha \in A^2 (\tilde{\mathcal{A}})$ is the map
\be
 \alpha: (p,q)\in\tilde{\mathcal{A}}\otimes\tilde{\mathcal{A}}\longrightarrow\tilde{\mathcal{A}}\label{coct}
\ee
with $\tilde{\mathcal{A}}$  the  algebra of Fourier transforms (to be precise $\alpha$   is defined on translations, realised as linear functions in $\tilde{\mathcal{A}}$)  and the coboundary operator
\be
\del: A^k (\tilde{ \mathcal{A}}) \longrightarrow A^{k+1} (\tilde{\mathcal{A}})
\ee
\bea
\del \gamma(p_0, ..., p_{k})&=&   \sum_{i=0}^{k}(-1)^{i}\gamma(p_0,...,p_{\hat i},p_{i+1},...,p_k)\nn\\
&-&(-1)^{k}\gamma(p_0-p_k, p_i-p_k,..., p_{k-1}-p_k)
\eea
In order for $\alpha$ in \eqn{coct} to be a  two-cocycle this becomes
\bea
0&=& \del \alpha(p,q,r)=\alpha(q,r) - \alpha(p,r) + \alpha(p,q) -\alpha(p-r,q-r)\nonumber\\
&=& 2\left(f\star(g\star h) -(f\star g) \star h\right)
\eea
that is \eqn{associativity}.
A straightforward calculation verifies that $\del^2=0$.
Thus,  the associativity
condition \eqn{associativity} is a 2-cocylce condition in the $\alpha$ cohomology. Analogously the commutativity condition can be shown to be a coboundary condition. Indeed, for $\alpha$ to be a coboundary it has to be
\be
\alpha(p,q)= \del\beta (p,q)= \beta(q)-\beta(p)+\beta(p-q)\label{cobt}
\ee
which implies the commutativity condition \eqn{commutativeprod}, that is $\alpha(p,q)= \alpha(p,p-q)$.

The Gr\"onewold-Moyal and Wick-Voros products, both
noncommutative, are respectively given by,
\begin{equation}
\alpha_{M}(p,q)=-\frac{\ii}{2}\theta^{ij}q_{\ii}(p_{j}-q_{j})=\frac{i}{2}\theta
p\wedge q
\end{equation}
and
\begin{equation}
\alpha_{V}(p,q)=-\theta
q_{-}(p_{+}-q_{+})=\alpha_{M}(p,q)-\frac{\theta}{2}(p-q)\cdot q
\end{equation}
which are both cocyles in the $\alpha$-cohomology and, more interestingly,
 differ by a term which is a $\alpha$-coboundary, according to \eqn{cobt} with $\beta$ so defined
 \be
 \beta(p)= p^2
 \ee
 Indeed we easily verify that
 \begin{equation}
\alpha_{V}(p,q)=\alpha_{M}(p,q)+\frac{\theta}{4}\del \beta(p,q)
\end{equation}

 With a
symbolic manipulation programme and a little work is not difficult
to construct viable polynomial $\alpha$'s. For example the
following expression in two dimensions gives rise to an
associative product:
\bea
\alpha&=&\gamma_1 p_2q_1+\gamma_2 p_1q_2
-(\gamma_1+\gamma_2)q_1q_2+\beta_1(p_2q_2^2-p_2^2q_2)\nonumber\\&&+\beta_2(\frac{p_2^2q_1-p_1q_2^2}2
+p_1p_2q_2-p_2q_1q_2)
\eea
for arbitrary $\gamma_1, \gamma_2, \beta_1$ and $ \beta_2$.

\subsection{The differential form of the product}

The second form is a generalization of the expression~\eqn{Moyalseries} and it is a series which depends on a ``small'' parameter which we call again $\theta$:
\be
f\star g=\sum_{r=0}^\infty C_r(f,g)\theta^r \ , \label{starexpans}
\ee
To recover the original commutative product in the
limit $\theta\to0$ we need to impose that
$C_0(f,g)=f g$. To ensure associativity
the remaining $C_r$'s have to satisfy the following properties,
\bea
f C_r(g,h)-C_r(f g,h)+C_r(f,g h)-C_r(f,g)
h\nonumber\\ =\sum_{j+k=r}
\left(C_j(C_k(f,g),h)-C_j(f,C_k(g,h))\right) \ , \label{Crcond}
\eea
for all $j,k,r>0$. The generalization to the multiparameter case is easily done considering $\theta$ and the $C$'s to have indices which are summed over.

A possible problem with this form of products is that it is
defined on the space of formal series in the coordinates, and
there is in general no control on the convergence of the series
after the product has been taken. Moreover not always the
differential form is useful for field theory. The quantity
$C_1(f,g)-C_1(g,f)$ gives a Poisson structure on the space which
is important for quantization. One defines the Poisson structure
as:
\be
\{f,g\}=C_1(f,g)-C_1(g,f)=\Lambda{ij}\del_i f \del_j g
\ee
where
\be
\Lambda^{ij}=\frac12\left(C_1(x^i,x^j)-C_1(x^j,x^i)\right)
\ee

Notice that if $C_1(f,g)=C_1(g,f)$ then the product is commutative. The proof is the following. First consider~\eqn{Crcond} for $r=2$ and $h=f=x^n$ and $g=x^m$. Then relation~\eqn{Crcond} becomes
\bea
f C_2(g,f)-C_2(f g,f)+C_2(f,g f)-C_2(f,g)
f\nonumber\\=x^n (C_2(x^m,x^n)-C_2(x^n,x^m))-C_2(x^{n+m},x^n)+C_2(x^n,x^{n+m})
\nonumber=\\ =
\left(C_1(C_1(f,g),f)-C_1(f,C_1(g,f))\right)=0 \label{c2cancel}
\eea
because of the symmetry of $C_1$. The second line of the above equation, has to hold for all $x$'s and therefore it must be $C_2(x^{n+m},x^n)=C_2(x^n,x^{n+m})$ for generic $n,m$. This implies that $C_2(f,g)=C_2(g,f)$. It is then possible to prove exactly in the same way that if $C_l(f,g)=C_l(g,f)$ for $l<r$ then all the terms in the r.h.s.\ of~\eqn{Crcond} pairwise cancel, and we are left to the equivalent of~\eqn{c2cancel} with a generic $r$, and then analogously prove that $C_r(f,g)=C_r(g,f)$.

Since the $c_i$'s are differential operators the product is
translationally invariant if and only if the $C$'s are
combinations of derivatives only. In this case
\be
\Lambda^{ij}=\frac12[x^i,x^j]_\star
\ee
There is a notion of equivalence which says that two star products are to be considered as equivalent if there exists a map $T$ such
that
\be
T(f)\star T(g)=T(f\star'g) . \label{equivalentprod}
\ee
According to this the Gr\"onewold-Moyal and Wick-Voros products are equivalent, the
map $T$ being given by $T=\e^{-\frac\theta4\nabla^2}$. A general
result of Kontsevich (in the context of formal series)
\cite{Kontsevich} proves that two products with the same Poisson
structure are equivalent. We have seen an instance of such an equivalence while calculating the UV behaviour of Feynman diagrams at one-loop. We have seen that, although the Green's functions are different for Moyal and Voros products, the UV behaviour is the same as well as the UV/IR mixing. We will see in the next section that this is a generic feature of translation invariant products and what counts is the cohomology class of $\alpha$ in the $\alpha$-cohomology.

\section{UV/IR Mixing for General Products}
\setcounter{equation}{0}

We are now ready to calculate the two point functions at one loop.
In this paper we are only interested to the ultraviolet properties
of the generalized products. The presence of the deformed product
also changes the propagator and it may alter the S-matrix. A full
analysis of a scattering process however requires to take into
account issues of symmetry, and the proper definition of the
incoming states. We have considered~\cite{paganini} the issue for
the case in which the product is coming from a twisted
symmetry~\cite{Wess,aletWess,Helsinki} (for a review
see~\cite{WSS}) and found that a proper treatment of the incoming
states and of the symmetries implies that there is no difference
between the Gr\"onewold-Moyal and Wick-Voros products.

We now proceed to the calculation of the loop contribution. We
first have to give the free propagator, which is
\begin{equation}\label{TPGF}
\tilde{G}^{2}_0(p)=\frac{e^{-\alpha(0,p)}}{p^{2}-m^{2}} \ .
\end{equation}
The presence of the exponential in the propagator alters its
properties. The analysis of this (free) propagator and its role in
the S-matrix involves a proper definition of the asymptotic states
and their normalization. Since in this paper we are only interested
in the corrections of the propagator due to the loop expansion,
and the ultraviolet/infrared mixing, we will not discuss this
issue. We just comment that in \cite{paganini} it is shown that in
the case of the Wick-Voros product, in the context of twisted
deformations, the exponential is absorbed in the normalization of
the in and out states.

In order to calculate the vertex, let us write down the
interacting term of the action in momentum space. Using the
definition of the product and the fact that the integral is a
trace we have
\begin{multline}
S_{\mathrm{int}}=\frac{g}{4!}
\int\dd x^d\,\dd k_{1}^d\,\dd k_{2}^d\,\dd k_{3}^d\,\dd k_{4}^d\,
\tilde{\phi}(k_{2})\tilde{\phi}(k_{1}-k_{2})\tilde{\phi}(k_{4})\tilde{\phi}(k_{3}-k_{4})\\
e^{\alpha(k_{1},k_{2})}e^{\alpha(k_{3},k_{4})}e^{ik_{1}\cdot x}\star e^{ik_{3}\cdot x}\\
=\frac{g}{4!}\int\dd k_{1}^d\,\dd k_{2}^d\,\dd k_{3}^d\,\dd k_{4}^d\,
\tilde{\phi}(k_{2})\tilde{\phi}(k_{1}-k_{2})\tilde{\phi}(k_{4})\tilde{\phi}(k_{3}-k_{4})\\
e^{\alpha(k_{1},k_{2})}e^{\alpha(k_{3},k_{4})}
\int\dd k^d\,e^{\alpha(0,k)}\delta(k_{1}-k)\delta(k_{3}+k).
\end{multline}
So
\begin{multline}
S_{\mathrm{int}}=\frac{g}{4!}
\int\dd k_{1}^d\,\dd k_{2}^d\,\dd k_{3}^d\,\dd k_{4}^d\,
\tilde{\phi}(k_{2})\tilde{\phi}(k_{1}-k_{2})\tilde{\phi}(k_{4})\tilde{\phi}(k_{3}-k_{4})\\
e^{\alpha(k_{1},k_{2})}e^{\alpha(k_{3},k_{4})}e^{\alpha(0,k_{1})}\delta(k_{1}+k_{3})\\
=\frac{g}{4!}\int\dd k_{1}^d\,\dd k_{2}^d\,\dd k_{3}^d\,\dd k_{4}^d\,
\tilde{\phi}(k_{2})\tilde{\phi}(k_{1}-k_{2})\tilde{\phi}(k_{4})\tilde{\phi}(k_{3}-k_{4})\\
e^{\alpha(k_{1},k_{2})+\alpha(k_{3},k_{4})+\alpha(0,k_{1})}\delta(k_{1}+k_{3})\\
=\frac{g}{4!}\int\dd k_{1}^d\,\dd k_{2}^d\,\dd k_{3}^d\,\dd k_{4}^d\,
\tilde{\phi}(k_{1})\tilde{\phi}(k_{2})\tilde{\phi}(k_{3})\tilde{\phi}(k_{4})\\
e^{\alpha(k_{1}+k_{2},k_{1})+\alpha(k_{3}+k_{4},k_{3})+\alpha(0,k_{1}+k_{2})}
\delta(k_{1}+k_{2}+k_{3}+k_{4}).
\end{multline}
Therefore the vertex is given by
\begin{equation}\label{V}
V_{\star}=V_0
e^{\alpha(k_{1}+k_{2},k_{1})+\alpha(k_{3}+k_{4},k_{3})+\alpha(0,k_{1}+k_{2})},
\end{equation}
where $V_0$ is the ordinary vertex defined in \eqn{V0def}. We now
proceed to the calculation of the four-point Green's function to
the tree level. To this end, we must attach to the vertex four
propagators. So we have up to a constant
\begin{align}
\tilde{G}^{(4)}=&e^{\alpha(k_{1}+k_{2},k_{1})+\alpha(k_{3}+k_{4},k_{3})+\alpha(0,k_{1}+k_{2})}
\prod_{a=1}^{4}\frac{e^{-\alpha(0,k_{a})}}{k_{a}^{2}-m^{2}}
\delta\left(\sum_{a=1}^{4}k_{a}\right)\\
=&\frac{e^{\alpha(k_{1}+k_{2},k_{1})+\alpha(k_{3}+k_{4},k_{3})+\alpha(0,k_{1}+k_{2})-\sum_{a=1}^{4}\alpha(0,k_{a})}}
{\prod_{a=1}^{4}(k_{a}^{2}-m^{2})}\delta\left(\sum_{a=1}^{4}k_{a}\right).
\end{align}
Consider now the two diagrams of figure~\ref{oneloop}. For the planar case (a) the correction is obtained using three propagators \eqref{TPGF},
one with momentum $p$, one with momentum $-p$, one with momentum
$q$ and the vertex \eqref{V} with assignments $k_{1}=-k_{4}=p$ and $k_{2}=-k_{3}=q$
and, of course, the integration in $q$. We have up to a constant
\begin{align}
\nonumber
G^{(2)}_{\rm P}=&\int\dd q^d\,\frac{\e^{-\alpha(0,p)-\alpha(0,-p)-\alpha(0,q)}}
{(p^{2}-m^{2})^{2}(q^{2}-m^{2})}\e^{\alpha(p+q,p)+\alpha(-p-q,-q)+\alpha(0,p+q)}\\
\nonumber
=&\int\dd q^d\,\frac{e^{-\alpha(0,p)-\alpha(0,-p)-\alpha(0,q)+\alpha(p+q,p)+\alpha(-p-q,-q)+\alpha(0,p+q)}}
{(p^{2}-m^{2})^{2}(q^{2}-m^{2})}\\
\nonumber
=&\int\dd q^d\,\frac{\e^{-2\alpha(0,p)-\alpha(0,q)+\alpha(p+q,p)+\alpha(-p-q,-q)+\alpha(0,p+q)}}
{(p^{2}-m^{2})^{2}(q^{2}-m^{2})}\\
=&\int\dd q^d\,\frac{\e^{-\alpha(0,p)}}
{(p^{2}-m^{2})^{2}(q^{2}-m^{2})}
\end{align}
where we used the last of \eqn{relations}. We see that with
respect to the commutative case the only correction is in the
factor $\e^{-\alpha(0,p)}$ which is the correction of the free
propagator. The ultraviolet divergences of the loop are the same
and therefore the short distance physics is unaffected (in this
aspect) by the star product. The correction of the free propagator
can then be reabsorbed in the S-matrix as done in \cite{paganini}.

Consider now the non-planar case in figure~\ref{oneloop}(b).
The structure is the same as in the planar case, but this time the assignments are
\begin{equation}
k_{1}=-k_{3}=p\quad\mathrm{and}\quad k_{2}=-k_{4}=q.
\end{equation}
We have up to a constant
\begin{align}
\nonumber
G^{(2)}_{\rm NP}=&\int\dd q^d\,\frac{\e^{-\alpha(0,p)-\alpha(0,-p)-\alpha(0,q)}}
{(p^{2}-m^{2})^{2}(q^{2}-m^{2})}\e^{\alpha(p+q,p)+\alpha(-p-q,-p)+\alpha(0,p+q)}\\
\nonumber
=&\int\dd q^d\,\frac{\e^{-\alpha(0,p)-\alpha(0,-p)-\alpha(0,q)+
\alpha(p+q,p)+\alpha(-p-q,-p)+\alpha(0,p+q)}}
{(p^{2}-m^{2})^{2}(q^{2}-m^{2})}\\
\nonumber
=&\int\dd q^d\,\frac{\e^{-2\alpha(0,p)-\alpha(0,q)+
\alpha(p+q,p)+\alpha(-p-q,-p)+\alpha(0,p+q)}}
{(p^{2}-m^{2})^{2}(q^{2}-m^{2})}\\
=&\int\dd q^d\,\frac{\e^{-\alpha(0,p)+\alpha(p+q,p)-\alpha(p+q,q)}}
{(p^{2}-m^{2})^{2}(q^{2}-m^{2})} \label{nonplanardiagram}
\end{align}
since by using again \eqref{IR} we have
\begin{align}
\nonumber
\alpha(-p-q,-p)&=-\alpha(0,-p-q)+\alpha(0,-p)+\alpha(0,-q)-\alpha(p+q,q)\\
&=-\alpha(0,p+q)+\alpha(0,p)+\alpha(0,q)-\alpha(p+q,q).
\end{align}
The one-loop corrections to the propagator in the non-planar case
can be rewritten as
\begin{equation}
G^{(2)}_{NP}=\int\dd q^d\,\frac{\e^{-\alpha(0,p)+\omega(p,q)}}
{(p^{2}-m^{2})^{2}(q^{2}-m^{2})},
\end{equation}
where we define
\begin{equation}
\omega(p,q)=\alpha(p+q,p)-\alpha(p+q,q).
\end{equation}
For the Gro\"onewold-Moyal product this term is the phase $\ii p_i\theta^{ij}q_j$.

This function has some useful property. First of all, it satisfies
the 2-cocycle condition \eqref{associativity}. Moreover,
\begin{align}
\omega(p,p)&=0\\
\omega(p,0)&=0\\
\omega(0,p)&=0\\
\omega(p,q)&=-\omega(q,p)&\quad\mathrm{antisymmetry}\\
\omega(-p,-q)&=\omega(p,q)&\quad\mathrm{parity} \label{omegaparity}\\
\omega(-p,q)&=\omega(p,-q)\\
\omega(p,q)&=\omega(p-q,p). \label{conditionuseful}
\end{align}
From \eqref{associativity} we have
\begin{equation}
\alpha(p+q,p)=\alpha(p+q,r)-\alpha(p,r)+\alpha(p+q-r,p-r)
\end{equation}
and by setting $r=q$ we get
\begin{equation}
\omega(p,q)=\alpha(p,p+q)-\alpha(p,q).
\end{equation}
This quantity vanishes if the product is commutative because of the condition~\eqn{commutativeprod}, that is $\omega$ is a 2-cocycle which is not a coboundary. This means that the nonplanar diagram captures the noncommutativity of the product, or, it only depends on the $\alpha$-cohomology class. In other words no change in the ultraviolet can come from a commutative product (an $\alpha$-coboundary).

We now prove that the contribution to the one loop diagram must necessarily be of the form $p_i\theta^{ij}q_j$ and that it depends on the Poisson structure induced by the star product. We will only need the rather mild assumption that $\alpha$ (and therefore $\omega$) can be Taylor expanded in a power series of $p$ and $q$. The parity relation~\eqn{omegaparity} requires the series to be composed only of even monomials. Let us express the function $\omega$ with a multi-index notation
\be
\omega(p,q)=\sum_{\vec i \,\vec j} a_{\vec i \,\vec j}\, p^{\vec i}q^{\vec j}
\ee
where $\vec i=(i_1,\ldots i_d)$ and
\be
p^{\vec i}=p_1^{i_1}p_2^{i_2}\ldots p_d^{i_d}
\ee
If we now use relation~\eqn{conditionuseful} we have that it must be
\be
\sum_{\vec i\, \vec j} a_{\vec i \,\vec j} \,q^{\vec i}(p^{\vec j}-(p-q)^{\vec j})=0
\label{azero}
\ee
this condition, because of the independence of $p$ and $q$ implies
that the coefficient $a$ must vanish \emph{except} in the case in
which all of the $j_a$'s but one vanish. In this case the
antisymmetry of the $a$'s ensures that~\eqn{azero} vanishes
without putting further constraints on the coefficient. Using
antisymmetry the same reasoning can be done for the first
multiindex and this shows that the term appearing in the one loop
amplitude is of the kind $\omega(p,q)=\ii\theta^{ij}p_iq_j$. Where
we added the imaginary unit to be consistent with the usual
notation. Using the relation \eqn{starcondition} is possible to
see that $\theta$ must be real.

In fact the expression which appears is the one  related to the
commutator of the coordinates. A straightforward calculation gives
\be
x^i\star x^j-x^j\star x^i=-\ii\frac{\partial\alpha}{\partial p_{i}}(0,0)x^{j}+
\ii\frac{\partial\alpha}{\partial p_{j}}(0,0)x^{i}-
\frac{\partial^{2}\alpha}{\partial p_{i}\partial q_{j}}(0,0)+
\frac{\partial^{2}\alpha}{\partial p_{j}\partial q_{i}}(0,0) \label{alphacomm}
\ee
The first two terms vanish because $\alpha$ has no linear term (we must justify this from associativity), while the second gives the coefficients of the quadratic terms in the expansion of $\alpha$ antisymmetrised. On the other side we have established that $\omega$ is quadratic as well and expressing
\be
\alpha(p,q)=\alpha_{ij}p^iq^j+\ldots
\ee
where $\ldots$ are terms cubic and above, we have
\be
\omega(p,q)=\ii\theta^{ij}p_iq_j=\alpha^{ij}(p_i+q_i)p_j-\alpha^{ij}(p_i+q_i)q_j=\alpha^{ij}(p_i+q_i)(p_j-q_j)
\ee
imposing the condition~\eqn{identitycondition} makes only the
mixed terms survive on the r.h.s., and the quadratic mixed terms
are the ones which appear in \eqn{alphacomm}.

We have shown that the term appearing in the exponent of the
nonplanar diagram \eqn{nonplanardiagram} is just the commutator of
the $x$'s multiplied by the external and internal momenta. The
Gr\"onewold-Moyal and Wick-Voros cases are therefore generic,
their behaviour is replicated by all translationally invariant
products.

Therefore we have shown that products with the same
Poisson structure (and hence the same commutator) which are  equivalent in the sense of Konsevitch,
have the same
structure of infrared/ukltraviolet mixing. We have also noted that
equivalence in the sense of~\eqn{equivalentprod} does not a priori
 mean physical equivalence. The propagators and
Green's functions are in general  different, as for  the
Gr\"onewold-Moyal and Wick-Voros products, where the Green's functions are not the same. In
this case however, using the fact that both products come from a
Drinfeld twist~\cite{Drinfeld}, and therefore have a deformed
symmetry by a quantum group~\cite{Wess,Helsinki,WSS}, it can be
shown~\cite{paganini} that the S-matrix is the same.

In fact it is easy to see that the general translational
invariant product \eqn{intprodtran}, in the case of $\alpha$
analytic comes from a Drinfeld twist. Expressing
\be
\alpha(p,q)=\sum{\vec i, \vec j}\alpha_{\vec i, \vec j}q^{\vec
i}(p-q)^{\vec j}
\ee
the product comes from the Drinfeld twist
\be
\mathcal{F}=\exp\left(-\sum_{\vec i, \vec j}\alpha_{\vec i, \vec
j}\del_x^{\vec i}\otimes\del_y ^{\vec j}\right)
\ee
then the product is
\bea
f\star g &=& \e^{\sum_{\vec i, \vec j}\alpha_{\vec i, \vec
j}\del_x^{\vec i}\del_y ^{\vec
j}f(x)g(y)\big|_{x=y}}\nonumber\\
\nonumber\\
&=& \int \dd p \dd q  \e^{\sum_{\vec i, \vec
j}\alpha_{\vec i, \vec j}p^{\vec i}q^{\vec j}}\tilde f(p) \tilde
g(q) \e^{\ii(x(p+q) )}
\eea
and the usual expression \eqn{intprodtran} is obtained with a
change of variables.

\section{Conclusions}
\setcounter{equation}{0}

In this paper we have shown, with an explicit calculation, that
the Ultraviolet/Infrared mixing found for the Gr\"onewold-Moyal
and Wick-Voros products is a generic feature of translationally
invariant associative star products. The vertex is changed by an
exponential which maintains invariance for cyclic permutation of
the external momenta but not for arbitrary exchanges. Therefore
the planar and nonplanar diagrams behave differently.

The planar diagrams have corrections to the propagator which are
unchanged with respect to the usual case. The nonplanar diagrams
on the other side present the phenomenon of Ultraviolet/Infrared
mixing. For high internal momentum the ultraviolet divergences are
damped by a phase, but these divergences reappear in the infrared
(low incoming momentum). The phase appearing in the exponent in
the nonplanar diagram is the one related to the commutator of the
coordinates. In a sense it may be said that the mixing is given
(in this translationally invariant case) by the Poisson structure
of the underlying space. This is non trivial, because the Green's
functions and the propagators of the theory are in general
different. What remains the same is the short distance behaviour.
Our calculation confirms the heuristic argument that the mixing is
a manifestation of the spacetime version of Heisenberg's
uncertainty, given by the Poisson structure.

\paragraph{Acknowledgments} One of us (FL) would like the Department of Estructura i Constituents de la materia, and the Institut de Ci\'encies del Cosmos,
Universitat de Barcelona for hospitality. His work has been supported in part by CUR Generalitat de
Catalunya under project 2009SGR502.

\end{document}